# ADVANCED SIMULATION TECHNIQUES FOR STRAYLIGHT PREDICTION OF HIGH PERFORMANCE MM-WAVE REFLECTING TELESCOPE

M. Sandri[(1)], F. Villa[(1)], M. Bersanelli[(2)], R.C. Butler[(1)], N. Mandolesi[(1)], J. Tauber[(3)]

on behalf of LFI consortium

[(1)]*IASF/CNR – Sezione di Bologna*
*Via P.Gobetti, 101*
*40129 – Bologna, Italy*
*Email: sandri@bo.iasf.cnr.it*

[(2)]*Università degli Studi di Milano*
*Via Celoria, 16*
*20133 – Milano, Italy*
*Email: marco@ifctr.mi.cnr.it*

[(3)]**ESA – ESTEC**
**PO Box 299**
**2200 AG Noordwijk, The Netherlands**
*Email: jtauber@rssd.esa.int*

## INTRODUCTION

The study of Cosmic Microwave Background (CMB) anisotropies represents one of the most powerful tools for Cosmology. The shape of the angular power spectrum of CMB anisotropies sensitively depends on fundamental cosmological parameters, so that an accurate measure of the spectrum provides a unique method to establish the parameters value with high precision. PLANCK, the third Medium-Size Mission (M3) of the European Space Agency, represents the third generation of mm-wave instruments designed for space observations of CMB anisotropies, following NASA's missions COBE and MAP. The PLANCK survey will cover the whole sky with unprecedented sensitivity ($\Delta T/T \sim 2 \div 4 \cdot 10^{-6}$), angular resolution (typically 10 arcmin at 100 GHz) and frequency coverage (from 30 to 857 GHz). The Low Frequency Instrument (LFI), operating in the $30 \div 100$ GHz range, is one of the two instruments onboard PLANCK, sharing the focal region of a 1.5 meter off-axis dual reflector telescope with the High Frequency Instrument (HFI) covering higher frequencies [1]. The LFI is an array of 46 HEMT-based pseudo-correlation receivers coupled to the telescope by 23 dual profiled corrugated feed horns working in four frequencies, with 12 feed horns at 100 GHz, 6 at 70 GHz, 3 at 44 GHz and 2 at 30 GHz. The layout of the focal assembly foresees the LFI feed horns distributed around the HFI, which is an array of 48 bolometric detectors, placed in the centre of the telescope focal surface, operating in six frequency channels between 100 and 857 GHz [2]. The wide frequency range covered by the combination of two instruments, will significantly improve the accuracy of the subtraction of foreground contamination from the primordial CMB anisotropy, providing at the same time a gold-mine of cosmological as well as astrophysical information [3].

Owing to the small level of the CMB anisotropies (about 10 μk rms), to reach these ambitious scientific goals, the control of systematic effects is mandatory. This requires a careful instrument design as well as an accurate knowledge of instrumental characteristics necessary to optimise the data analysis phase. In this framework, we consider here the effect of "Straylight" on the LFI detectors, i.e. unwanted off-axis radiation contributing to the observed signal. More precisely, Straylight is defined for PLANCK instruments as the radiative power that reaches a detector within its RF bandwidth, and does not originate from sources in the main beam (i.e., at angles θ less than or about 1°). Straylight induces a signal that is indistinguishable from those induced by sources located in the main beam and its variations are source of systematic error [4]. Therefore, the behaviour of the PLANCK antenna, both at intermediate (θ between 1° and 5°) and large angles (θ greater than 5°) from the directions of the telescope Line of Sight (LOS), have to be carefully considered. The requirement for the Straylight rejection does not pertain only to the telescope itself, but rather to the coupling between feeds and the external environment, the latter composed by all the satellite surfaces (including telescope) and celestial objects which can emit, reflect, or scatter radiation within the operational bandwidth of the detectors [5]. In particular, the PLANCK optical environment is dominated by a large baffle surrounding the telescope and, on the bottom, the higher of the three thermal shields (V-grooves) used to thermally decouple the telescope and focal assembly from the service module (for more details, see Villa et al., these proceeding). The antenna response features at large angles from the beam centre (far side lobes) are determined largely by diffraction and scattering from the edges of the mirrors and from nearby supporting structures. Therefore they can be reduced by decreasing the illumination at the edge of the primary mirror, i.e. by increasing the Edge Taper (ET), defined as the ratio of the power per unit area incident on the centre of the mirror to that incident on the edge. Of course, the higher is the ET, the lower is the side lobe level and the Straylight contamination. On the other hand, increasing the ET has a negative impact on the angular resolution, which reduces the ability to reconstruct the anisotropy power spectrum of the CMB at high multipoles. In order to obtain the best angular resolution achievable compatible with Straylight requirements, a trade-off study must be carried out. By means of accurate radiation pattern simulations, the details of the antenna response for each LFI feed horn are calculated and the feed horn design can be optimised. In this work we present simulation techniques based on advanced GTD methods that we are using to evaluate the Straylight rejection of the entire PLANCK/LFI optical system in a rigorous way, even at very low levels.

## SIMULATION METHODS

The simulations are performed by considering the feed as a source and by computing the pattern scattered by both reflectors on the far field using GRASP8, a software developed by TICRA (Copenhagen, DK) for analysing general reflector antennas. To predict the radiation pattern, different techniques can be applied: Physical Optics (PO), Physical Theory of Diffraction (PTD), Geometrical Optics (GO), and Geometrical Theory of Diffraction (GTD) [6]. PO is the most accurate method and may be used in all regions of the space surrounding the reflector antenna system. The field of the source is propagated on the reflector to calculate the current distribution on the surface. Then, the currents are used for evaluating the radiated field from the reflector. The calculation of the currents close to the edge of the scatterer are modelled by PTD. The radiated PTD field is a correction which has been added to the PO field to obtain the scattered field. Unfortunately, as the frequency increases the reflectors have to be more and more precisely sampled. As a consequence, the density of the integration grids, in which currents have to be computed on the reflectors, must be finer and the computation time can become huge. For a two-reflectors antenna system like PLANCK, the computation time increases with the fourth power of the frequency. Without considering the effect of the shields, this means more than 800 hours to perform a full pattern at 100 GHz, using a 550 MHz Pentium II machine.

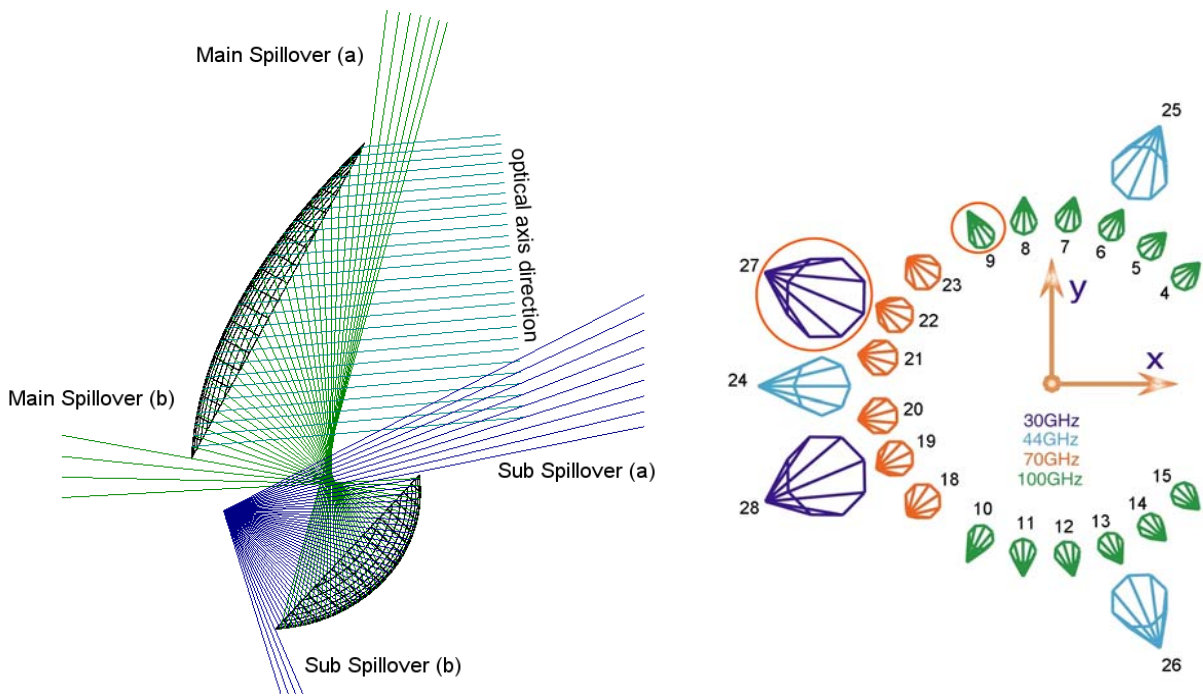

Fig. 1 *Left side* – Sketch of the PLANCK Telescope in the symmetry plane: the Main and Sub Reflector Spillover are shown. The rays marked with (b) will be blocked and redirected by shields. *Right side* – PLANCK /LFI Focal Plane Unit configuration represented in the Reference Detector Plane axis system (RDP). The two feed horn under study (LFI 27 at 30 GHz and LFI9 at 100 GHz) are identified with a circle.

Moreover, proper evaluation of the effect of the shield is crucial for the optimisation of the edge taper, since the shields redistribute the power that is radiated by the horns and is not reflected by the telescope. In principle, to predict accurately the pattern including the shields, a full PO computation of the whole spacecraft geometry would be required. However, this would be exceedingly time consuming and not applicable in practice. For this reason, in our simulations we use a new GRASP8 add-on package. This package, named Multi-reflector GTD (MrGTD), computes GTD fields from any number of reflectors sequentially illuminated starting at a given source and represents a suitable method for predicting the full-sky radiation pattern of complex mm-wavelength optical systems [7]. Reflector geometry, source characteristics, and output field points have to be defined, together with each contribution (i.e., a bundle of rays defined by a sequence of scatterers and by the type of interaction – reflection or diffraction – occurred on each of them) to be taken into account to reach an accurate radiation pattern prediction. Although MrGTD is in general less time consuming than PO (Physical Optics), it must be applied in a rigorous way in order to obtain truthful results, specially at low power levels (down to ~ -50dBi). Considering that it is a new simulation method, we carried out a dedicated study in order to reach confidence and good understanding of the method for a reliable application to the PLANCK case.

**PATTERN PREDICTION FOR PLANCK LFI**

As mentioned, MrGTD computes the scattered field from the selected reflectors performing a backward ray tracing. For many scatterers, the amount of ray tracing contributions may lead to unacceptable computation times. The main problem is to understand the contributions that we have to consider, or in other words, to identify the sequence of diffractions and/or reflections on each scatterer which produce a significant power level in the resulting radiation pattern. For example, considering the two reflectors, the baffle and the first V-groove as blocking structures, the simplest (first order) contributions producing significant power levels are reflections on the sub reflector (Rs, where "R" means reflection and "s" means sub reflector), on the main reflector (Rm, where "R" means reflection and "m" means main reflector), and on the baffle (Rb, where "R" means reflection and "b" means baffle), as well as diffractions on the sub reflector (Ds, where "D" means diffraction and "s" means sub reflector), on the main reflector (Dm, where "D" means diffraction and "m" means main reflector), and on the baffle (Db, where "D" means diffraction and "b" means baffle). Other non-negligible contributions can be found considering two interactions with the reflectors (second order – for example, rays reflected on the baffle and then diffracted by the main reflector: RbDm), three interactions (third order – for example, rays reflected on the sub reflector, diffracted by the main reflector, and then diffracted by the baffle: RsDmDb) and so on. We have not considered reflections or diffractions on the V-groove because we expect these contributions at very low levels. In order to select all and only the significant contributions, our team has developed a software that automatically recognizes those contributions for which the maximum power level is greater than a given threshold, which we set to about -100 dB (-50 dBi at 30 GHz and -40 dBi at 100 GHz) [8]. In the following, full pattern simulations considering up to two interactions are presented. An evaluation of higher order contributions is currently in progress for several feed models, for the 30 and 100 GHz LFI channels.

**PATTERN WITHOUT SHIELDS**

Antenna patterns (without considering the effect of the shields) have been computed at 30 and 100 GHz, in spherical field grids with $\theta \in [-180°,180°]$ and $\varphi \in [0°,180°]$ ($\Delta\theta = 0.5°$ and $\Delta\varphi = 0.5°$), according to the GRASP8 definition , in the main beam coordinate system of the considered feed horn. The main beam region has been simulated using PO/PTD in order to avoid caustics artefacts originated by the GTD approach.

**LFI27 at 30 GHZ**

The feed horn considered in these simulations (30 GHz LFI feed horn #27) is specified by its spherical wave expansion provided by Alcatel Space Industries, since the sub reflector is in the near field of the corrugated horn and near field effects cannot be neglected. The feed horn directivity is about 22 dBi and the ET is 30 dB at 22°. The location of this horn is shown in Fig. 1 and the relative main beam in the sky is about 4° from the LOS. At 30 GHz, where the computational time is not huge, a comparison with a full PO analysis has been performed. The results are shown in Fig. 2, for the cut at $\varphi = 45°$. The contributions considered in the final simulation are, at the first order: direct, Rs, Rm, Ds, Dm, and, at the second order: DsDm, RsDm, DsRm, DmDm, RmDm, DmRm, DsRs, DsDs. Differences in cuts computed with the two methods are probably due to some contributions not yet considered in the MrGTD analysis. Six contributions with three interactions between the surfaces have been identified in the regions where differences appear: DsDsRm, DsRsRm, RsDmDs, DsRsDm, DmRmDm, and RsDmRm.

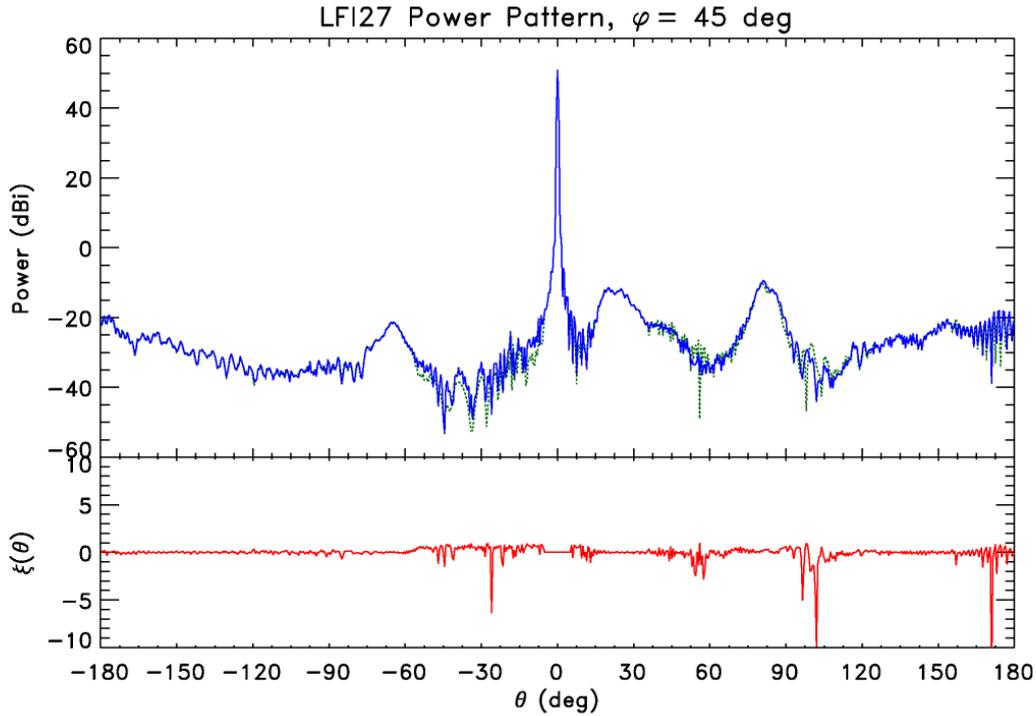

Fig. 2 Comparison between full PO analysis (solid line) and MrGTD (dashed line) for PLANCK/LFI Feed Horn #27 at 30 GHz. Cut at $\varphi = 45°$ is shown. The main spillover peak covers a region near $\varphi = 25°$ and $\theta = 80°$, with a power level of about -10 dBi. $\xi(\theta)$ is defined as $[P_{PO/PTD}(\theta) - P_{MrGTD}(\theta)] / P_{PO/PTD}(\theta)$ and represents the difference between the two curves. In this cut, the baffle would block the radiation coming roughly from $\theta < 0°$ and redirect it in the intermediate beam region and near the Main Spillover.

**LFI9 at 100 GHz**

The feed model used in these electromagnetic simulations is a dual profiled one for the 100 GHz LFI feed horn #9. The location of this horn is shown in Fig. 1 and the relative main beam in the sky is about 3° from the LOS. Three different patterns have been computed by Modal Matching/MoM Models on dual profile corrugated horns (by the CAISMI Institute, Florence – Italy) in 72 azimuthally equidistant cuts, in which $\theta$ ranges from 0° to 180°. The first represents the current LFI9 Qualification Model (PG25, ET 25.5 dB at 24°). The Edge Taper of the other two horns has been chosen by means of optical simulations based on gaussian feed horns, in order to improve the resulting angular resolutions on the sky. Two different designs have been performed, PG27 (9A) and PG31 (9B), that lead to angular resolutions of 10 and 9.5 arcmin, respectively. The contributions considered in the final simulation are those in which the maximum power level computed by the program is greater than -40 dBi (-100 dB at 100 GHz). To first order, the contributions from direct, Rs, Rm, Ds, Dm, have been selected and, to second order the contributions from RsDm, DsRm, DsDm, DsRs, DmRs. The results are shown in Fig. 3 for the cuts at $\varphi = 45°$. It should be noted how the main spillover, at about $\varphi = 45°$ and $\theta = 90°$, raises increasing the illumination at the edge of the primary mirror (-7 dBi with an ET 25.5 at 24°, -2 dBi with an ET 19.0 at 24°, and 3 dBi with an ET 15.0 dB at 24°) since it is due to diffraction and scattering from the edges of the mirror. In Fig. 4 the pattern over the full solid angle for the LFI9 PG27 is shown. The boresight direction is at the top of the map. The $\varphi$ angle goes from 180° (left side of the map) to 0° (centre of the map), which corresponds to 180°. Then $\varphi$ decreases from 180° to 0° again (right side of the map). Each polar cut with a constant $\varphi$ value is a meridian of the map. The $\theta$ angle runs, on each meridian, from 0° (towards the main beam direction) to 180° (-180°) sweeping the map on its left (right) side. Therefore the centre of the map has $\varphi$ and $\theta$ values equal to 0° and 90°, respectively. In Fig. 4 the major contributions are pointed out. The Main Reflector Spillover is mainly due to reflection on sub reflector not intercepted by the primary mirror (Rs), diffraction on the main reflector (Dm), and diffraction on the sub reflector (Ds). The Sub Reflector Spillover is mainly due to the rays from feed which pass the secondary mirror (direct) and diffraction on the sub reflector (Ds). Another important contribution is caused by the feed side lobe (direct). Cuts reported in Fig. 3 can be seen in Fig. 4 in a different representation and the peaks of the Main and Sub Reflector Spillover can be easily identified.

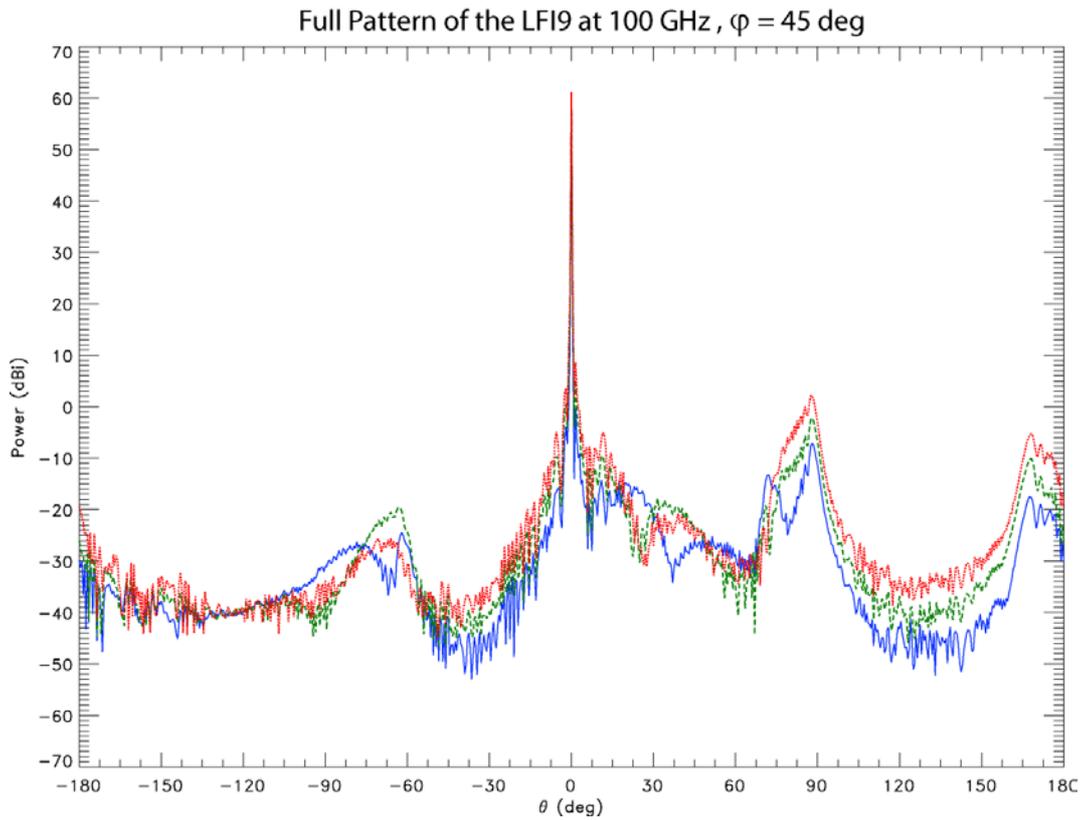

Fig. 3 Full Pattern of the feed horn # 9 at 100 GHz. Three curves with different ET are shown for the cut with φ = 45°: 25.5 dB at 24° (solid line), 19 dB at 24° (dashed line), and 15 dB at 24° (dotted line).

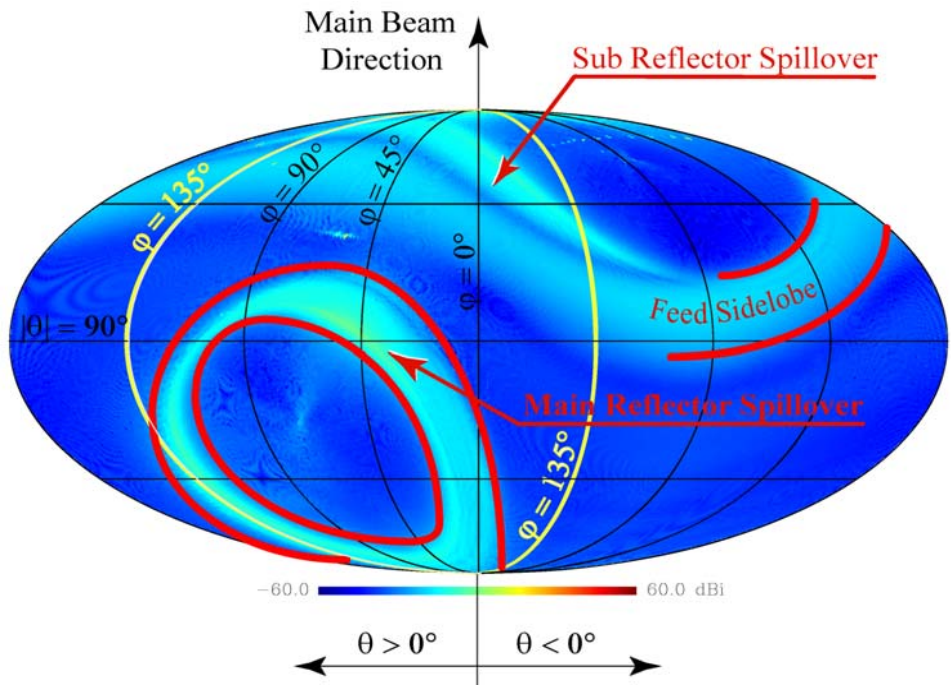

Fig. 4 Example of a Full Pattern computed using GRASP8 MrGTD and plotted with HEALPix [9]. In the map, major contributions are pointed out, together with an indication of φ and θ angles (LFI9).

**PATTERN WITH SHIELDS**

Feed Horn #27 and #9 reported in the previous section have been simulated taking into account the effect of the shields. The blocking structures considered in this study consist of the baffle and the upper V-groove, which is directly in view of the telescope and instrument (see Fig. 6).

**LFI27 at 30 GHz**

Contributions for MrGTD analysis of the feed horn #27 (threshold level -50 dBi) are, at the first order: direct, Rs, Rm, Rb, Ds, Dm, Db, and, at the second order: RsDm, DsRm, DbRm, RbDb, DsDm, DbDm, DsDb, RbDm, RbDs, DmRm, DmDm, RmDm, RsRb, RbRm, RbRs, DsRb,DmRb, RsDb, DmDb, DsDs, RmRb, RmDb, DsRs, DbDb. An evaluation of the third order contributions is currently in progress and about fifty contributions have already been found. In Fig. 5 some contributions are shown ($\Delta\theta = 0.5°$ and $\Delta\varphi = 0.5°$). Each map is computed adding the new contribution to the previous one. It should be noted that: 1) the contribution due to the rays reflected by the sub reflector and not intercepted by shields, covers the Main Spillover region raising the power level up to -4 dBi, 2) two wings appear at about $\varphi \approx 70°$ and $\varphi \approx 160°$ ($\theta \approx 50°$) with a power level up to -2 dBi (see the $2^{nd}$ map), 3) the intermediate main beam region is made dirty by the diffractions on the main reflector (see the $6^{th}$ map). Fig. 7 shows the map computed considering all contributions with a power level greater than -50 dBi with up 2 interactions. This preliminary analysis shows that the maximum side lobe level is about -9 dBi (-59 dB from the main beam power peak, at $\theta \sim 85°$ and $\varphi \sim 22°$). This is in line with the LFI requirements for Straylight.

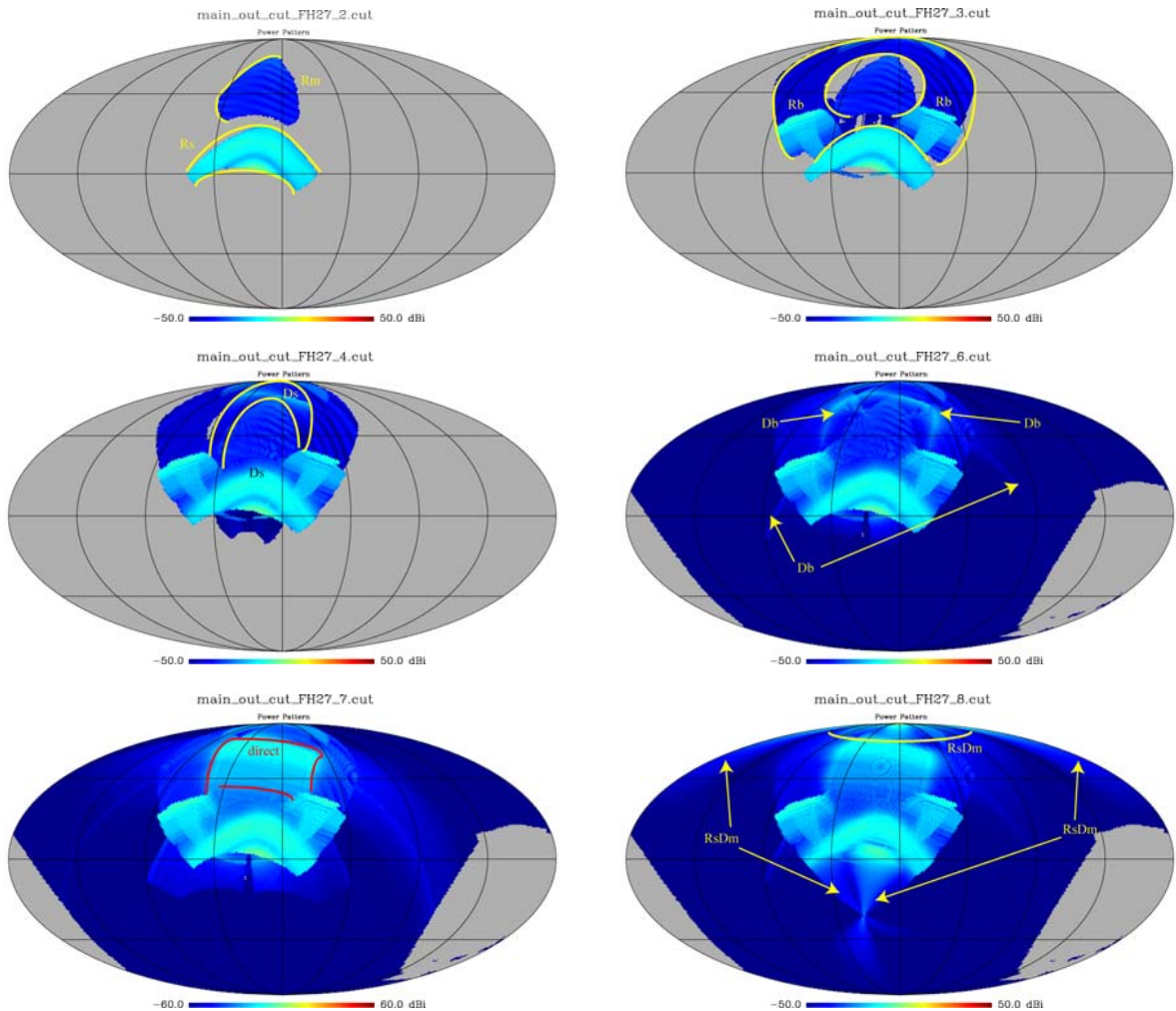

Fig. 5 Some contributions for the Multi-Reflector GTD simulation of LFI Feed Horn #27 at 30 GHz.

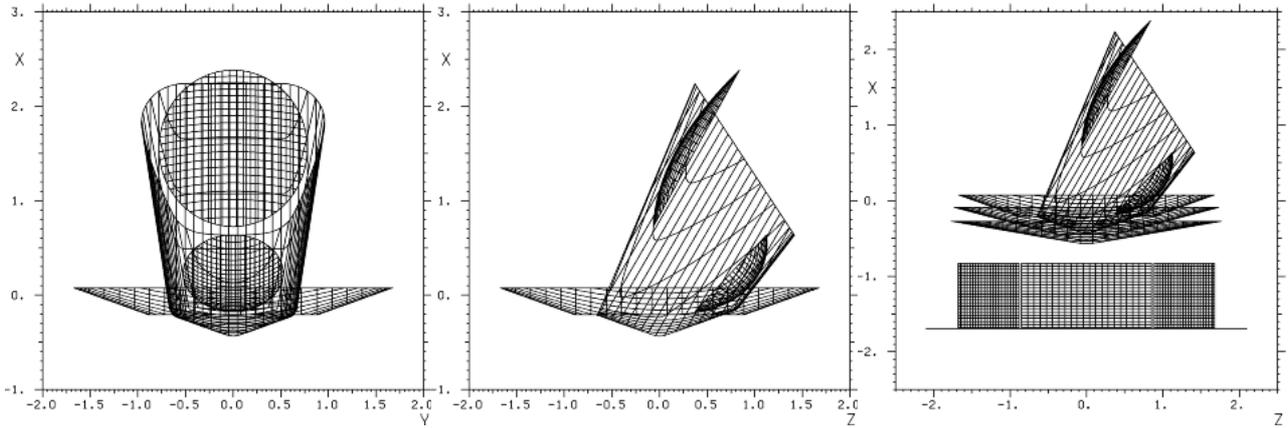

Fig. 6 Telescope and shields geometry in (XY)$_{TEL}$ (*left*) and (XZ)$_{TEL}$ (*centre*) planes; spacecraft geometry (*right*).

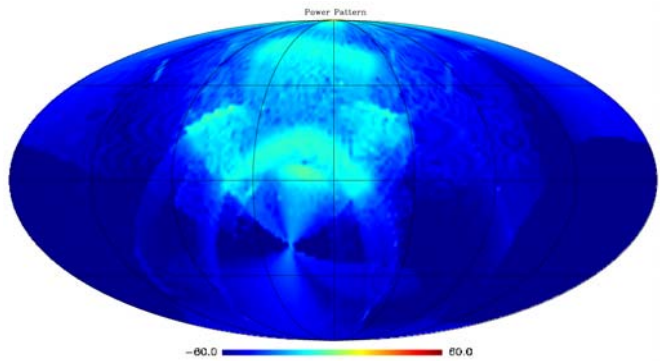

Fig. 7 Map of the LFI Feed Horn #27 at 30 GHz computed with MrGTD considering all contributions with a power level greater than -50 dBi with up 2 interactions.

**LFI9 at 100 GHz**

Contributions for MrGTD analysis of the LFI Feed Horn 9 PG25 at 100 GHz (threshold level: -40 dBi) are, to first order: direct, Rs, Rm, Rb, Ds, Dm, Db, and, at the second order: DbRm, RsRb, DsRm, DsDm, RsRm, RsDm, RbRm, RbDm, RsDb, RbDb, DsRb, DsDb, RbDs, RmRb, DmRb. Twenty contributions have been found to be significant at the third-order level, with power greater than -40 dBi, and their computation is currently in progress. In Fig. 8 the map computed considering all contributions with a power level greater than -40 dBi with up 2 interactions is shown. This preliminary analysis shows that the maximum side lobe level is about -6 dBi (-67 dB from the main beam power peak, at $\theta \sim 85°$ and $\varphi \sim 50°$). This is in line with the LFI requirements for Straylight.

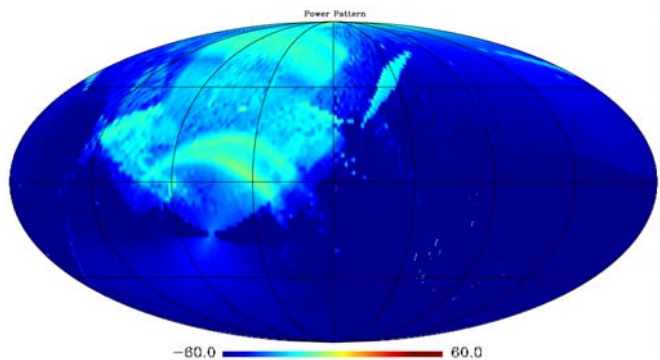

Fig. 8 Map of the LFI Feed Horn #9 PG25 at 100 GHz computed with MrGTD considering all contributions with a power level greater than -40 dBi with up 2 interactions.

# CONCLUSIONS

Straylight contamination may be one of the most critical sources of systematic effects in observations of CMB anisotropies by future satellite missions like PLANCK. While accurate measurements of the antenna pattern will be necessary for a firm evaluation of this effect, optical simulations are of primary importance for an optimal knowledge of instrumental characteristics, especially in the far side lobes where the power levels are extremely low and make direct measurements difficult and uncertain. To predict the radiation pattern, Physical Optics represents the most accurate method, but at millimetre wavelength the size of the reflectors and the other structures may be hundreds of wavelengths and this approach is no longer applicable. In the framework of the PLANCK Low Frequency Instrument optical interface optimisation, the shield contribution is crucial in electromagnetic simulations aimed to define the edge taper of each LFI feed horn since shields redistribute the power radiated by horns that is not reflected by the telescope. GRASP8 Multi-Reflector GTD is an advanced GTD method for evaluating the Straylight rejection of the entire optical system, including shields, much less time consuming than PO. It computes GTD fields from any number of reflectors sequentially illuminated starting at a given source. Reflector geometry, source characteristics, and output field points have to be defined, together with all the relevant contributions. We have developed a dedicated software tool in order to optimise the critical process of selecting *all and only* the contributions exceeding a given threshold specified in terms of a power level. The two maps reported in Fig. 7 and Fig. 8 have been computed considering all contributions with a power level greater than -50 dBi and -40 dBi, respectively, with up 2 interactions. This preliminary analysis, which is based on a representative subset of the LFI feeds, shows that the maximum side lobe levels are about -59 dB and -67 dB, for the 30 and 100 GHz feed horns respectively, which is compatible with the LFI requirements for the Straylight. Simulations of third order contributions are currently in progress.

# ACKNOWLEDGMENTS

We wish to thank people of the Herschel/Planck Project, ALCATEL Space Industries, and the LFI Consortium that are involved in activities related to optical simulations. The HEALPix package use is acknowledged (see HEALPix home page at http://www.eso.org/science/healpix/). Thanks are also due to Per Nielsen (TICRA Engineering Consultants, Copenhagen), for his clarifying correspondence about Multi Reflector GTD and support with GRASP8.